\documentstyle[aps,epsf,floats]{revtex}
\begin{document}
\author{Yan V. Fyodorov$\P,\S$ and
Hans-J\"{u}rgen Sommers$\S$ }

\title{Spectra of Random Contractions and Scattering Theory for
Discrete-Time Systems}
\address{$\P$ Department of Mathematical Sciences,
Brunel University, Cleveland
Road, Uxbridge, UB8 3PH, UK  }
\address{$\S$Fachbereich Physik, Universit\"at-GH Essen,
D-45117 Essen, Germany        }
\date{\today}
\maketitle
\begin{abstract}

 Random contractions (sub-unitary random matrices) appear
naturally when considering quantized chaotic maps within
a general theory of open linear stationary systems
with discrete time. We analyze statistical properties of
complex eigenvalues of generic $N\times N$ random matrices $\hat{A}$
of such a type, corresponding to systems with broken time-reversal invariance.  
Deviations from unitarity are characterized by
rank $M\le N$ and a set of eigenvalues $0<T_i\le 1, i=1,...,M$
of the matrix $\hat{T}=\hat{{\bf 1}}-\hat{A}^{\dagger}\hat{A}$.
We solve the problem completely by deriving
the joint probability density of $N$ complex eigenvalues and calculating all
$n-$ point correlation functions.
In the limit $N>>M,n$ the correlation functions
acquire the universal form found earlier
for weakly non-Hermitian random matrices.
\end{abstract}
\vfill

The theory of wave scattering can be looked at as an integral part
of the general theory of linear dynamic open systems in terms of
the input-output approach. These ideas and relations
 were developed in system theory and
engineering mathematics many years ago, see papers \cite{Liv,Ar,Hel}
and references therein. Unfortunately, that
 development went almost unnoticed by the majority of physicists
 working in the theory of chaotic quantum scattering and related
phenomena, see \cite{LW,FS} and references therein.
For this reason we feel it could be useful to
recall some basic facts of the input-output approach in
such a context.

An Open Linear System is characterized by three Hilbert spaces:
 the space $E_0$ of internal states $\Psi\in E_0$
and two spaces $E_{\pm}$ of incoming (-) and outgoing (+)
 signals or waves also called input and output spaces,
made of vectors $\phi_{\pm}\in E_{\pm}$. Acting in these three spaces
are four operators, or matrices: a) the so-called
fundamental operator $\hat{A}$ which maps any vector from internal
space $E_0$ onto some vector from the same space $E_0$ b)
two operators $\hat{W}_{1,2}$, with $\hat{W}_1$ mapping incoming
states onto an internal state and $\hat{W}_2$ mapping internal states onto
outgoing states and c) an operator $\hat{S}_0$ acting from $E_{-}$ to
$E_{+}$.

We will be interested in describing
the dynamics $\Psi(t)$ of an internal state with time $t$
provided we know the state at
initial instant $t=0$  and
the system is subject to a given input signal $\phi_{-}(t)$.
In what follows we consider only the case of
 the so-called stationary (or {\it time-invariant}) systems
when the operators are assumed to be time-independent.
Let us begin with the case of continuous-time description.
The requirements of linearity, causality and
stationarity lead to a system of two dynamical equations:
\begin{equation}\label{1}
\begin{array}{c}
i\frac{d}{d t}\Psi=\hat{A} \Psi(t)+\hat{W}_1\phi_{-}(t)\\
\phi_{+}(t)=\hat{S}_0\phi_{-}(t)+i\hat{W}_2\Psi(t)
\end{array}
\end{equation}
Interpretation of these equations depends on the nature of the state vector
$\Psi$ as well as of the vectors $\phi_{\pm}$ and is different in
different applications.
In the context of quantum mechanics one
relates the scalar product $\Psi^{\dagger}\Psi$ with the probability to find  
a particle inside the "inner" region at time $t$, whereas  
$\phi^{\dagger}_{\pm}\phi_{\pm}$ stays for probability currents flowing in  
and out of the region of internal states (the number of particles coming
or leaving the inner domain per unit time). The condition of particle  
conservation then reads as:
\begin{equation}\label{2}
\frac{d}{dt}\Psi^{\dagger}\Psi=
\phi^{\dagger}_{-}\phi_{-}-\phi^{\dagger}_{+}\phi_{+}
\end{equation}
It is easy to verify that Eq.(\ref{2})
 is compatible with the dynamics Eq.(\ref{1}) only provided the
operators satisfy the following relations:
\[
\hat{A}^{\dagger}-\hat{A}=i\hat{W}\hat{W}^{\dagger}\,\,,\,\,  
\hat{S}_0^{\dagger}\hat{S}_0=\hat{{\bf 1}}\ \, \mbox{and}\,
\hat{ W}^{\dagger}\equiv \hat{W}_2=-\hat{S}_0\hat{W}_1^{\dagger}
\]
which shows, in particular, that $\hat{A}$ can be written as  
$A=\hat{H}-\frac{i}{2}\hat{W}\hat{W}^{\dagger}$, with a Hermitian  
$\hat{H}=\hat{H}^{\dagger}$.

The meaning of $\hat{H}$ is transparent:
it governs the evolution $i\frac{d}{d t}\Psi=\hat{H} \Psi(t)$ of an
inner state $\Psi$ when the
coupling $\hat{W}$ between the inner space
and input/output spaces is absent. As such, it is
just the Hamiltonian describing
the closed inner region. The fundamental operator
$\hat{A}$ then has a natural
interpretation of the effective non-selfadjoint Hamiltionian describing
the decay of the probability from the inner region at
zero input signal: $\phi_{-}(t)=0$ for any $t\ge 0$.
If, however, the input signal is
given in the Fourier-domain by $\phi_{-}(\omega)$,
the output signal is related to it by:
\begin{equation}\label{3}
\phi_{+}(\omega)=\left[\hat{S}(\omega)\hat{S}_0\right]\phi_{-}(\omega)\quad,\quad  
\hat{S}(\omega)=
\hat{{\bf 1}}-i\hat{W}^{\dagger}\frac{1}{\omega \hat{{\bf 1}}-\hat{A}}\hat{W}
\end{equation}
where we assumed $\Psi(t=0)=0$. The unitary matrix $\hat{S}(\omega)$ is  
known in the mathematical
literature as the characteristic matrix-function of the non-Hermitian operator
$\hat{A}$. In the present context it is just the scattering matrix
whose unitarity is guaranteed by the conservation law Eq(\ref{2}).

The contact with the theory of chaotic scattering is
now apparent: the expression Eq.(\ref{3}) was
frequently used in the physical literature
\cite{LW,FS} as a starting point for extracting universal
properties of the scattering matrix for a quantum chaotic system
 within the so-called random matrix approach.
The main idea underlying such an approach
is to replace the actual Hamiltonian
$\hat{H}$ by a large random matrix and to
calculate the ensuing statistics of the scattering matrix.
The physical arguments in favor of such a replacement can be found
in the cited literature.

In particular, most recently the
statistical properties of complex eigenvalues of
the operator $\hat{A}$ as well as related quantities
were studied in much detail\cite{FS,FK,SFT,Frahm}.
Those eigenvalues are poles of the scattering matrix and
have the physical interpretation of
{\it resonances} - long-lived intermediate states to which discrete
energy levels of the closed system are transformed due
to coupling to continua.

In the theory presented above the time $t$ was a continuous parameter.
On the other hand, a very useful instrument in the analysis of classical
Hamiltonian systems with chaotic dynamics are the so-called area-preserving
 chaotic maps\cite{Sm2}. They appear naturally either as a mapping
of the Poincar\'{e} section onto itself, or as result of a stroboscopic
description of Hamiltonians which are periodic functions of time.
Their quantum mechanical analogues are unitary operators which act on
Hilbert spaces of finite large dimension $N$. They are often referred
to as evolution, scattering or Floquet operators, depending on the
physical context where they are used. Their eigenvalues consist of
$N$ points on the unit circle (eigenphases). Numerical studies of
various classically chaotic systems suggest that the eigenphases
conform statistically quite accurately the results obtained for
unitary random matrices of a particular symmetry (Dyson circular ensembles).

Let us now imagine that a system represented by
 a chaotic map ("inner world") is embedded in a
larger physical system ("outer world") in such a way
that it describes particles which can
come inside the region of chaotic motion and leave
it after some time. Models of such type appeared, for example, in \cite{BGS}
where a kind of scattering theory for "open quantum maps" was developed
based on a variant of Lipmann-Schwinger equation.

On the other hand, in the general system theory \cite{Ar,Hel}
dynamical systems with discrete time are considered as frequently as
those with continuous time.
 For linear systems  a "stroboscopic"  dynamics is just a linear map  
$(\phi_{-}(n); \Psi(n))\to(\phi_{+}(n); \Psi(n+1))$ which can be generally  
written as:
\begin{equation}\label{5}
\left(
\begin{array}{c}
\Psi(n+1)\\ \phi_{+}(n)
\end{array}
\right)
=\hat{V}
\left(\begin{array}{c}
 \Psi(n)\\ \phi_{-}(n)\end{array}\right)\quad,\quad
\hat{V}=\left(\begin{array}{cc}
\hat{A}&\hat{W}_1\\\hat{W}_2&\hat{S}_0\end{array}\right)
\end{equation}
Again, we would like to consider a conservative system, and the
discrete-time analogue of Eq.(\ref{2}) is:
\[
\Psi^{\dagger}(n+1)\Psi(n+1)-\Psi^{\dagger}(n)\Psi(n)=
\phi^{\dagger}_{-}(n)\phi_{-}(n)-\phi^{\dagger}_{+}(n)\phi_{+}(n)
\]
which amounts to unitarity of the matrix $\hat{V}$ in Eq.(\ref{5}).
In view of such a unitarity $\hat{V}$
of the type entering Eq.(\ref{5}) can always be parametrized as (cf.\cite{Mel}):
\begin{equation}  \label{6}
\hat{V}= \left(\begin{array}{cc} \hat{u}_1&0\\0&\hat{v}_1\end{array}\right)
\left(\begin{array}{cc} \sqrt{1-\hat{\tau}\hat{\tau}^{\dagger}}& -\hat{\tau}
\\\hat{\tau}^{\dagger}&\sqrt{1-\hat{\tau}^{\dagger}\hat{\tau}}
\end{array}\right)\left(\begin{array}{cc}
\hat{u}_2&0\\0&\hat{v}_2\end{array}\right)
\end{equation}
where the matrices $u_{1,2}$ and $v_{1,2}$ are unitary and
$\hat{\tau}$ is a rectangular $N\times M$ diagonal matrix with the entries
$\tau_{ij}=\delta_{ij}\tau_j, \, 1\le i\le N\,,\,  1\le j\le M\quad 0\le  
\,\tau_j\le 1$.

Actually, it is frequently convenient to redefine
input, output and internal state
as: $\phi_{-}(n)\to \hat{v}_2^{-1}\phi_{-}(n),\quad \phi_{+}\to \hat{v}_1
\phi_{+}(n)$ and $\Psi(n)\to\hat{u}_2\Psi(n)$ which amounts just to
 choosing an appropriate basis in the corresponding spaces.
The transformations bring the matrix $\hat{V}$ to a somewhat simplier
form:
\begin{equation}  \label{7}
\hat{V}=
\left(\begin{array}{cc} \hat{u} \sqrt{1-\hat{\tau}\hat{\tau}^{\dagger}}& -u  
\hat{\tau}
\\ \hat{\tau}^{\dagger}&\sqrt{1-\hat{\tau}^{\dagger}\hat{\tau}}\end{array}
\right)
\end{equation}
where $ \hat{u}= \hat{u}_2^{\dagger}\hat{u}_1$. Such a form suggests
a clear interpretation of the constituents of the model. Indeed,
for $\hat{\tau}=0$ the dynamics of the system amounts to:
$\Psi(n+1)=\hat{u}\Psi(n)$. We therefore identify $\hat{u}$ as a unitary
evolution operator of the "closed" inner state domain decoupled both
from input and output spaces. Correspondingly,
$\hat{\tau}\ne 0$ just provides a coupling that makes the system
open and converts the fundamental operator $\hat{A}=\hat{u}
\sqrt{1-\hat{\tau}\hat{\tau}^{\dagger}}$ to a contraction:
$1-\hat{A}^{\dagger}\hat{A}=\tau\tau^{\dagger}\ge 0$. As a result,
the equation $\Psi(n+1)=\hat{A}\Psi(n)$ describes an irreversible
decay of any initial state $\Psi(0)$ for zero input $\phi_{-}(n)=0$,
whereas for a nonzero input and $\Psi(0)=0$ the Fourier-transforms
$\phi_{\pm}(\omega)=\sum_{n=0}^{\infty}e^{in\omega}\phi_{\pm}(n)$
 are related by a unitary scattering matrix $\hat{S}(\omega)$
given by:
\begin{equation} \label{8}
\hat{S}(\omega)=\sqrt{1-\hat{\tau}^{\dagger}\hat{\tau}}
-\hat{\tau}^{\dagger}\frac{1}{e^{-i\omega}-\hat{A}}\hat{u}\hat{\tau}
\end{equation}

Assuming further that the motion outside the inner region is regular,
we should be able to describe generic features
of open quantized chaotic maps
 choosing the matrix $\hat{u}$ to be a member
of a Dyson circular ensemble.
Then one finds: $\hat{\tau}^{\dagger}\hat{\tau}=1-\left|\overline{\hat{S}
(\omega)}\right|^2$ , with the bar standing for the averaging of  
$\hat{S}(\omega)$ in Eq.(\ref{8}) over $\hat{u}$.
 Comparing this result with \cite{LW,FS} we see that
the $M$ eigenvalues $0\le T_i\le 1$ of the $M\times M$ matrix  
$\hat{\tau}^{\dagger}\hat{\tau}$ play the role of the so-called transmission  
coefficients and describe a
particular way the chaotic region is coupled to the outer world.

In fact, this line of reasoning is motivated by recent papers
\cite{Kol,KS}. The authors of \cite{Kol} considered the Floquet
description of a Bloch particle in a constant force and periodic
driving. After some approximations the evolution of the system is
described by a mapping: ${\bf c}_{n+1}={\bf F}{\bf c}_{n}$, where
the unitary Floquet operator ${\bf F}=\hat{S}\hat{U}$ is the product of a
unitary "M-shift"
$\hat{S}: S_{kl}=\delta_{l,k-M}, \, l,k=-\infty,...,\infty$
and a unitary matrix $\hat{U}$. The latter is
effectively of the form
$\hat{U}=\mbox{diag}(\hat{d_1},\hat{u},d_2)$, where $\hat{d}_{1,2}$
are (semi)infinite diagonal matrices
 and $\hat{u}$ can be
taken from the ensemble of random $N\times N$ unitary matrices.

One can check that such a dynamics can be easily brought to the standard
Eqs.(\ref{5},\ref{6}) with the fundamental operator being an $N\times N$
random matrix of the form
$\hat{A}=\sqrt{{\bf 1}-\hat{\tau}^{\dagger}\hat{\tau}} \hat{u}$, and all
$M$ diagonal elements of the $N\times M$ matrix $\tau$ are equal to
unity. Actually, the original paper \cite{Kol} employed a slightly
different but equivalent construction dealing with
an "enlarged" internal space of the dimension $N+M$. We prefer to
follow the general scheme because of its conceptual clarity.

Direct inspection immediately shows that the non-vanishing
 eigenvalues of the fundamental operator $\hat{A}$ as above coincide
with those of a $(N-M)\times (N-M)$ subblock of the  random
unitary matrix $u$. Complex eigenvalues of such "truncations" of random
unitary matrices were studied in much detail by the authors of
a recent insightful paper \cite{KS}. They managed to study eigenvalue
correlations analytically for arbitrary $N,M$. In particular,
they found that in the limit $N\to \infty$ for fixed $M$ these
correlation functions practically coincide \cite{Kol} with those obtained
earlier \cite{FS,FK} for operators of the form $\hat{A}=\hat{H}-
\frac{i}{2}\hat{W}\hat{W}^{\dagger}$ occuring in the theory of
open systems with continuous-time dynamics.

Such a remarkable universality, though not completely unexpected,
deserves to be studied in more detail.
In fact, truncated unitary
matrices represent only a particular case of random
contractions $\hat{A}$.  Actually, some statistical properties
of general subunitary matrices  were under investigation recently
as a model of scattering matrix for systems with absorption,
see \cite{abs}.

The particular case of rank-one deviations from unitarity
is the simplest one to investigate and was considered
in a recent preprint\cite{prel}. However, generalization to
arbitrary $M$ along the lines of \cite{prel} seemed to be problematic.
The main goal of the present paper
is to suggest a regular way of studying the specta of random contractions
for a given deviation from unitarity.

The ensemble of general $N\times N$ random contractions
$\hat{A}=\hat{u}
\sqrt{1-\hat{\tau}\hat{\tau}^{\dagger}}$ describing the chaotic map with broken
time-reversal symmetry
can be described by the following probability measure in the matrix space:
\begin{equation}\label{0}
{\cal P}(\hat{A})
d\hat{A}\propto \delta(\hat{A}^{\dagger}\hat{A}-\hat{G}) d\hat{A}\quad, \quad
\hat{G}\equiv 1-\hat{\tau}\hat{\tau}^{\dagger}
\end{equation}
where $d\hat{A}=\prod_{ij} d\hat{A}_{ij}dA^{*}_{ij}$ and we assumed that the  
unitary matrix $\hat{u}$ is taken from the Dyson circular unitary ensemble.
The $N\times N$ matrix $\hat{\tau}\hat{\tau}^{\dagger}={\bf 1}-
\hat{G}\ge 0$  is natural to call the deviation matrix
and we denote it $\hat{T}$.
It has $M$ nonzero eigenvalues coinciding
 with the transmission coefficients $T_a$ introduced above.
The particular choice $T_{i\le M}=1,\,\, T_{i> M}=0$
corresponds to the case considered in \cite{KS}.
In what follows we assume all $T_i<1$, but the resulting
expressions turn out to be valid in the limiting case $T_i=1$ as well.

Our first step is, following \cite{FK,KS,prel},
introduce the Schur decomposition
$\hat{A}=\hat{U}(\hat{Z}+\hat{R})\hat{U}^{\dagger}$
of the matrix $A$ in terms of a unitary $\hat{U}$,
diagonal matrix of the eigenvalues $\hat{Z}$ and a lower triangular
$\hat{R}$. One can satisfy oneself, that
the eigenvalues $z_1,...,z_N$ are generically not degenerate, provided
all $T_i<1$. Then, the measure written in
terms of new variables is given by
$d\hat{A}=|\Delta(\{z\})|^2 d\hat{R}d\hat{Z}d\mu(U)$, where
the first factor is just the Vandermonde determinant of eigenvalues $z_i$
and $d\mu(U)$ is the invariant measure on the unitary group.
The joint probability density of complex eigenvalues is then given by:
\begin{eqnarray}\label{9}
{\cal P}(\{z\})&\propto& |\Delta(\{z\})|^2 \int d\mu(U)d\hat{R}\
\delta\left((\hat{Z}+\hat{R})(\hat{Z}+\hat{R})^{\dagger}-
\hat{U}^{\dagger}\hat{G}\hat{U}\right)
\end{eqnarray}
The integration over $\hat{R}$ can be performed with some manipulations
using its triangularity (some useful hints can be found in \cite{KS}).
As the result, we arrive at:
\begin{eqnarray}\label{10}
{\cal P}(\{z\})\propto |\Delta(\{z\})|^2
\int d\mu(U) \prod_{l=1}^N
\delta\left(|z_1|^2...|z_l|^2-
\det\left[1-\hat{T}\hat{U}\hat{P}_l\hat{U}^{\dagger}\right]\right)
\end{eqnarray}
where $\hat{P}=\mbox{diag}
\left(1,...1,0,...,0\right)$, with first $l$ entries being equal to
unity, is a projector.

To perform the remaining integration over the unitary group
was mentioned as the main technical problem
in \cite{prel}, and proved to be more difficult than
the corresponding procedure for non-Hermitian matrices, see
\cite{FK}. Here we outline the main steps and present some
 intermediate expressions relegating details of the calculation
to a more extended publication.
First, one considers the columns of $N\times N$ unitary matrices as
$N-$ component of (mutually orthogonal) vectors $\,{\bf a}_l,\
l=1,...,N$ and introduces $N\times N$  matrices of rank $l$:
$\hat{Q}_l=\sum_{i=1}^l{\bf a}_i\otimes {\bf a}^{\dagger}_i$ ,
so that $\hat{U}\hat{P}_l \hat{U}^{\dagger}=\hat{Q}_l$.
Writing down the corresponding constraints in a form of
$\delta-$ functions, one can represent the
expression Eq.(\ref{9}) in the form:
\begin{eqnarray}\label{11}
{\cal P}(\{z\})\propto |\Delta(\{z\})|^2
\int \left(\prod_{l=1}^{N-1} d\hat{Q}_l\right) \prod_{l=1}^N
\delta\left(|z_1|^2...|z_l|^2-
\det(1-\hat{T}\hat{Q}_l\right)\prod_{i=1}^N\int d{\bf a}d{\bf
a}^{\dagger} \delta\left(\hat{Q}_i-\hat{Q}_{i-1}-
{\bf a}\otimes{\bf a}^{\dagger}\right)
\end{eqnarray}
where the matrices $\hat{Q}_l$ are considered to be unconstrained
$N\times N$ Hermitian. We also used the orthonormality condition
$\hat{Q}_N=1$ as well as the convention $\hat{Q}_0=0$.

Due to the fact that only $M$ out of $N$ eigenvalues
of the matrix $\hat{T}$ are non-zero, both the matrices $\hat{Q}_l$ and
the vectors ${\bf a}$ can be effectively taken to be of the
size $M$ ( it amounts to changing the unspecified normalisation
constant in Eq.(\ref{11})) and redefine the matrix $\hat{T}$
as $\hat{T}=\mbox{diag}(T_1,...,T_M)=\hat{\tau}^{\dagger}\hat{\tau}$.
Then it is convenient to change: $\hat{T}^{1/2}\hat{Q}_l\hat{T}^{1/2}\to  
\hat{Q}_l$ and separate integration over eigenvalues and eigenvectors of
matrices $\hat{Q}_l$. The latter can be performed in a recursive way
$l\to l+1$, with the multiple use of the  
Itzykson-Zuber-Harish-Chandra\cite{IZHC} formula.
After quite an elaborate manipulation, one finally arrives at the
following representation:
\begin{eqnarray}\label{main}
{\cal P}(\{z\})&\propto& \frac{\det^{M-N}(\hat{T})}{\det(1-\hat{T})
\prod_{c_1<c_2}\left(T_{c_1}-T_{c_2}\right)}
\prod_{c_1<c_2}\left(\frac{\partial}{\partial \tau_{c_1}}
-\frac{\partial}{\partial \tau_{c_2}}\right)
\int d\hat{\lambda}
e^{-i\mbox{Tr}{\hat{\tau}\hat{\lambda}}}|\Delta(\{z\})|^2
\prod_{k=1}^Nf(\ln{|z_k|^2},\hat{\lambda}),
\end{eqnarray}
where we defined the diagonal matrices of size $M$ as:
$\hat{\tau}=\mbox{diag}(\tau_1,...,\tau_M)\,,\,
\hat{\lambda}=\mbox{diag}(\lambda_1,...,\lambda_M)$
and used the notations: $\tau_c=\ln{(1-T_c)}$ and
\begin{equation}
f(a,\hat{\lambda})=i^{M-1}\sum_{q=1}^M\frac{e^{i\lambda_qa}}
{\prod_{s(\ne q)}(\lambda_q-\lambda_s)}\quad.
\end{equation}

The distribution Eq.(\ref{main}) is written for $|z_k|^2\le 1$ for
any $k=1,...,N$ and vanishes otherwise. The remarkable feature of
such a distribution is that it allows for calculation of all
$n-$point correlation functions for arbitary $N,n,M$ with help of
 the method of orthogonal polynomials. Again, the particular case
 $M=1$ \cite{prel} is quite instructive and can be recommended to follow  
first for understanding of the general formulae outlined below.

To this end, we write
\begin{eqnarray}
|\Delta(\{z\})|^2
\prod_{k=1}^N f(\ln{|z_k|^2},\hat{\lambda})=
\prod_{k=1}^N N_k(\hat{\lambda})\det{\left[\sum_{n=1}^N
\frac{(z_iz_j^*)^{n-1}}{N_n(\hat{\lambda})}
f(\ln{|z_j|^2},\hat{\lambda})\right]_{i,j=1,...N}}\quad.
\end{eqnarray}
where the constants $N_n(\hat{\lambda})$ are provided by the orthonormality
condition:
\begin{equation}
\int_{|z|^2\le 1}d^2z z^{m-1}(z^*)^{n-1}f(\ln{|z|^2},\hat{\lambda})
=\delta_{m,n}N_n(\hat{\lambda})\quad,
\end{equation}
which yields after a simple calculation $N_n(\hat{\lambda})=\pi\prod_{c=1}^M
\frac{1}{(n+i\lambda_c)}$.

Now, by applying the standard machinery of orthogonal polynomials
\cite{Mehta} one can find the correlation function:
\begin{eqnarray}
R_n( z_1,...,z_n)=\frac{N!}{(N-n)!}\int d^2z_{n+1}...d^2z_N{\cal
P}(\{z\}) \quad
\end{eqnarray}
as given by:
\begin{eqnarray}\label{genres}
R_n( z_1,...,z_n)\propto \hat{{\cal D}}
\int d\hat{\lambda}
e^{-i\mbox{Tr}{\hat{\tau}\hat{\lambda}}}
\prod_{k=1}^N N_k(\hat{\lambda})
\det{\left[K(z_i,z_j;\hat{\lambda})\right]}_{(i,j)=1,...,n} \quad,
\end{eqnarray}
where the kernel $K$ is defined as:
\begin{equation}
K(z_1,z_2;\hat{\lambda})=\frac{1}{\pi}
\sum_{n=1}^N\det{(i\hat{\lambda}+n)}(z_1z_2^*)^{n-1}
f(\ln{|z_2|^2},\hat{\lambda})
\end{equation}
and the differential operator $\hat{\cal D}$ is just the expression
in front of the $\lambda-$ integral in Eq.(\ref{main}).

In principle, all $\lambda-$ integrations in the equation
Eq.(\ref{genres}) can be performed explicitly and the resulting
 formulae provide the desired general solution of the problem.
However, for arbitary $N,M,n$ the results obtained in that way
are still quite cumbersome. We present below as an example
the lowest correlation
function $R_1(z)$, which is just the mean eigenvalue density inside
the unit circle $|z|<1$. It can be calculated from the following
recursive relation connecting the density for $M$ and $M-1$ open
channels:
\begin{equation}\label{den}
R_1^{(M)}(z)=R_1^{(M-1)}(z)+\frac{1}{\pi}\frac{\partial}{\partial
|z|^2} {\cal F}_1^{(M)}\{T_c;|z|^2\}{\cal
F}_2^{(M-1)}\{T_c;|z|^2\} \quad,
\end{equation}
where
\begin{eqnarray}
{\cal F}_1^{(M)}\{T_c;|z|^2\}&=&\sum_{c=1}^M
\left[1-\left(\frac{1}{|z|^2}-1\right)
\left(\frac{1}{T_c}-1\right)\right]^{N-1}
\frac{\theta(|z|^2-1+T_c)}{\prod_{s\ne c}\left(\frac{1}{T_s}-\frac{1}
{T_c}\right)}\quad,\\
{\cal F}_2^{(M-1)}\{T_c;|z|^2\}&=&\frac{|z|^{2N}}{(N-1)!}
\int_0^{\infty} dt e^{-t|z|^2}t^{N-1}\prod_{c=1}^{M-1}
\left(\frac{1}{T_c}-1+\frac{1}{t}\frac{\partial}{\partial |z|^2}\right)
\frac{1-|z|^{2N}}{1-|z|^2}\quad .
\end{eqnarray}

For the case of all equivalent channels, i.e. when all the transmission
coefficients $T_c$ are equal: $T_c=T$, such a recursive relation
can be represented in a more compact form:
\begin{equation}
R_1^{(M)}(z)=R_1^{(M-1)}(z)+\frac{1}{(M-1)!}\left(\frac{\partial }
{\partial t}\right)^{M-1}{\cal R}(t,z)|_{t=0}\quad,
\end{equation}
where the generating function ${\cal R}(t,z)$ is given by:
\begin{equation}
{\cal R}(t,z)=\frac{1}{\pi}\frac{\partial}{\partial |z|^2}
\frac{\xi^N-\eta^N}{\xi-\eta}
\end{equation}
and
\begin{equation}
\xi=1+(t-1)\left(\frac{1}{|z|^2}-1\right)
\left(\frac{1}{T}-1\right)\quad,\quad
\eta=1+(t-1)\left(1-|z|^2\right)\frac{1}{T}\quad.
\end{equation}

All the equations above are valid for arbitrary $N\ge M,n$.
 In the theory of quantum chaotic scattering we, however,
expect a kind of universality in the semiclassical limit. Translated
to the random matrix language such a limit corresponds
to $N\to \infty$ at fixed $n,M$.  Still, extracting the asymptotic behaviour
of the correlation function $R_n(z_1,...,z_n)$ from Eq.(\ref{genres})
in such a limit is not a completely straightforward task.
A useful trick is to notice that Eq.(\ref{genres}) can be rewritten
as:
\begin{eqnarray}\label{gr1}
R_k( z_1,...,z_n)&\propto&
\sum_{q_1=1,...,q_n=1}^M F_{q_1,...,q_n}\left(\{T_c;z\}\right)\quad,\\
F_{q_1,...,q_n}\left(\{T_c;z\}\right)&=&{\cal B}\{T_c\}
\det{\left[\sum_{k=1}^N\left(\frac{\partial}{\partial \tau_1}+k\right)
... \left(\frac{\partial}{\partial
\tau_M}+k\right)(z_iz_j^*)^{k-1}\right]}_{i,j=1,...,n}\\
\nonumber &\times&\prod_{c_1< c_2}
\left(\frac{\partial}{\partial\tau_{c_1}}
-\frac{\partial}{\partial \tau_{c_2}}\right)
 \int \prod_c\left(d\lambda_c
\frac{\exp{-i\lambda_c\tau_c}}{\prod_{l=1}^N(l+i\lambda_c)}\right)
\frac{e^{-i\sum_{j=1}^n\lambda_{q_j}\ln|z_j|^2}}
{\prod_{j=1}^n\prod_{s\ne q_j}^M(\lambda_{q_j}-\lambda_s)}  \quad,
\end{eqnarray}
where we introduced the notation:
\[ {\cal B}\{T_c\}=
\frac{1}{\prod_{c_1<c_2}\left(T_{c_1}-T_{c_2}\right)}
\prod_{c=1}^M\frac{T_c^{M-N}}{(1-T_c)}\quad.
\]
Introducing now the auxiliary differential operator
$\hat{{\cal D}}_{q_1,...,q_n}=\prod_{j=1}^n\prod_{s\ne q_j}^M
\left(\frac{\partial}{\partial \tau_{q_j}}-
\frac{\partial}{\partial \tau_s}\right)$
and considering its action upon the ratio $F_{q_1,...,q_n}/
{\cal B}\{T_c\}$ one can satisfy oneself that in the limit $N\gg M,n$
the leading contribution to $F_{q_1,...,q_n}$ is given by:
\begin{eqnarray}\label{interm}
F_{q_1,...,q_n}&\propto& \prod_{c=1}^M\theta(1-\tilde{T}_c)
\frac{(1-\tilde{T}_c)}{(1-T_c)}
\left(\frac{\tilde{T}_c}{T_c}\right)^{N-M}
\prod_{j=1}^n\prod_{s\ne q_j}^M
\left(\frac{1}{ T_{q_j}}-
\frac{1}{T_s}\right)^{-1}
\det{\left[K(z_i,z_j;\{\tilde{T_c}\})\right]_{(i,j)=1,...,n}}\quad,
\end{eqnarray}
where the kernel is given by
\begin{eqnarray}\label{kern}
K(z_i,z_j;\{\tilde{T_c}\})&=&
\sum_{k=1}^N\prod_{c=1}^M\left[(N-M)\frac{1-\tilde{T}_c}
{\tilde{T}_c}+k-1\right](z_iz_j^*)^{k-1}\\
\nonumber &=& \prod_{c=1}^M\left[(N-M)\frac{1-\tilde{T}_c}
{\tilde{T}_c}+x\frac{d}{dx}\right]\frac{1-x^N}{1-x}\left.\right|_{x=z_1z_2^*}  
\end{eqnarray}
where we used the notation:
$\tilde{T}_c=1-\exp{\left(\tau_c-
\sum_{j=1}^n\delta_{q_j,c}\ln{|z_j|^2}\right)}$.

Further simplifications occur after taking into account that
eigenvalues $z_i$ are, in fact,  concentrated
typically at distances  of order of $1/N$ from the unit circle.
Then it is natural to introduce new variables
$y_i,\phi_i$ according to $z_i=(1-y_i/N) e^{i\phi_i}$
and consider $y_i$ to be of the order of unity when $N\to \infty$.
First of all, one immediately finds that:
\begin{equation}\label{rat}
\lim_{N\to \infty}
\prod_{c=1}^M\left(\frac{\tilde{T}_c}{T_c}\right)^{N-M}
=\exp{\left[-2\sum_{j=1}^n y_j\frac{1-T_{q_{j}}}{T_{q_{j}}}\right]}
\end{equation}
 As to the phases $\phi_i$, we expect their typical separation scaling as:
$\phi_i-\phi_j=O(1/N)$.
 Now it is straightforward to perform explicitly the limit
$N\to \infty$ in Eq.(\ref{kern}). Combining all factors together, one
brings the correlation function
Eq.(\ref{gr1}) to the final form:
\begin{eqnarray}\label{fin}
R_n(z_1,...,z_n )&\propto&
\prod_{k=1}^n\sum_{q=1}^M
\frac{e^{-g_q y_k}}{\prod_{s\ne q}(g_q-g_s)}
\det{\left[\int_{-1}^{1}d\lambda \prod_{c=1}^M(\lambda+g_c)
e^{-\frac{i}{2}\lambda\delta_{ij}}\right]_{i,j=1,n}}
\end{eqnarray}
with $g_c=2/T_c-1$ and $\delta_{ij}=N(\phi_i-\phi_j)-i(y_i+y_j)$.
The expression above coincides in every detail with that obtained
in\cite{FK} for random GUE matrices deformed by a finite rank anti-Hermitian
perturbation.\footnote{One should remember that the mean
density of phases $\phi_i$ along the unit circle is $\nu=1/(2\pi)$
and take into account that the constants $g_c$ defined in \cite{FK}
are, in fact, $\pi\nu g_c=g_c/2$ in the notations of the present paper.}.
This completes the proof of universality for finite-rank deviations.

YVF is obliged to B. Khoruzhenko for discussions and suggestions
on the earlier stage of the research.
The financial support by SFB 237 "Unordnung und grosse Fluktuationen"
as well as of the grant No. INTAS 97-1342 is acknowledged with thanks.


\begin{thebibliography}{99}

\bibitem{Liv} M.S.Livsic "Operators, Oscillations, Waves: Open
Systems." (Translations of Mathematical Monographs, v.34 (AMS,
Providence, RI, 1973)
\bibitem{Ar} D.Z.Arov {\it Sib.Math.Journ.} {\bf 20} (1979), 149
\bibitem{Hel} J.W.Helton {\it J.Funct.Anal.} {\bf 16} (1974), 15
\bibitem{LW} C.H.Lewenkopf and H.A.Weidenm\"{u}ller {\it Ann.Phys. NY}
{\bf 212} (1991), 53
\bibitem{FS} Y.V.Fyodorov and H.-J.Sommers {\it J.Math.Phys.} {\bf
38} (1997), 1918
\bibitem{FK} Y.V.Fyodorov and B.A.Khoruzhenko {\it Phys.Rev.Lett.}
{\bf 83} (1999), 65
\bibitem{SFT} H.-J.Sommers, Y.V.Fyodorov and M.Titov {\it J.Phys.A} {\bf 32}
(1999), L77
\bibitem{Frahm} K.Frahm et al., {\it Europhys.Lett.}, {\bf 49}, (2000) 48 ;
H.Schomerus et al., {\it Physica A} {\bf 278} (2000) 469
\bibitem{Sm2} U.Smilansky in: {\it Supersymmetry and Trace Formulae:
Chaos and Disorder}, edited by I.V.Lerner et al., (Kluwer, NY, 1999),
p. 173
\bibitem{BGS} F.Borgonovi,~I. Guarneri and D.L.Shepelyansky
{\it Phys.Rev.} {\bf A 43}, 4517 (1991); F.Borgonovi and I.Guarneri
{\it Phys.Rev.} {\bf E 48} R2347 (1993)
\bibitem{Mel} P.A.Mello and J.-L.Pichard {\it J.de Phys. I} {\bf 1},
493 (1991)
\bibitem{Kol} M.Gl\"{u}ck, A.R.Kolovski and H.J.Korsch
{\it Phys.Rev.} {\bf E 60}, 247 (1999)
\bibitem{KS} K.Zyczkowski and H.-J. Sommers {\it J. Phys. A: Math.
Gen.} {\bf 33} (2000) 2045
\bibitem{abs} E. Kogan, P.A.Mello and He Liqun {\it Phys.Rev.}
{\bf E 61}( 2000) R17; C.W.J. Beenakker and P.W.Brouwer
{\it e-preprint} cond-mat/9908325
\bibitem{prel} Y.V.Fyodorov e-preprint {\it nlin.CD/0002034} at
xxx.lanl.gov
\bibitem{IZHC} C.Itzykson and J.B.Zuber {\it J.Math.Phys}, {\bf 21},
411 (1980); Harish-Chandra {\it Am.J.Math.} {\bf 79}, 87 (1957)
\bibitem{Mehta} M.L. Mehta {\it Random Matrices} 2nd ed. (Academic
Press, London, 1991)


\end{thebibliography}
\end{document}